\documentclass[aps,pre,twocolumn,floatfix,superscriptaddress,showpacs]{revtex4-1}

\usepackage{graphicx,amsmath,amssymb,color,epsf,verbatim}

\makeatletter
\AtBeginDocument{\@ifpackageloaded{natbib}{\ifNAT@numbers\if@filesw\immediate\write\@auxout{\string\global\string\NAT@numberstrue}\fi\fi}{}}
\makeatother
\begin{document}

\setcounter{page}{1}
\title{Branching process approach for Boolean bipartite networks of metabolic reactions}
\author{Deokjae Lee}
\affiliation{Department of Physics and Astronomy, Seoul National University, Seoul 151-747, Korea}
\author{K.-I. Goh}
\email{kgoh@korea.ac.kr}
\affiliation{Department of Physics, Korea University, Seoul 136-713, Korea}
\author{B. Kahng}
\email{bkahng@snu.ac.kr}
\affiliation{Department of Physics and Astronomy, Seoul National University, Seoul 151-747, Korea}
\date{\today}

\begin{abstract}
The branching process (BP) approach has been successful in explaining the avalanche dynamics in complex networks.
However, its applications are mainly focused on unipartite networks, in which all nodes are of the same type.
Here, motivated by a need to understand avalanche dynamics in metabolic networks, we extend the BP approach to a particular bipartite network composed of Boolean AND and OR logic gates.
We reduce the bipartite network into a unipartite network by integrating out OR gates, and obtain the effective branching ratio for the remaining AND gates. Then the standard BP approach is applied to the reduced network, and
the avalanche size distribution is obtained. We test the BP results with simulations on the model networks and two microbial metabolic networks, demonstrating the usefulness of the BP approach.
\end{abstract}


\maketitle

{\it Introduction---} The multiplicative branching process (BP) approach~\cite{harris} has been successful in helping us understand a variety of physical phenomena in complex networks such as percolation~\cite{havlin}, epidemic spreading~\cite{epidemic}, and avalanche dynamics~\cite{goh}. Such an approach is valid when these physical phenomena do not form a nontrivial fraction of loop structures in the process of forming clusters, spreading diseases, and toppling cascades.  Indeed, during the dynamic process in complex networks, the formation of loops is a rarity, and thus, the BP approach has been regarded as a useful method.  

Recent studies applying the BP approach to percolation, epidemic spreading, and avalanche dynamics were limited to unipartite networks, in which all nodes are of the same type. However, there are many examples of bipartite networks in real-world networks, in which the nodes belong to one of two types. The BP approach has also
been applied to such bipartite networks~\cite{newman2001} and other multitype networks~\cite{vazquez,allard} to study
the formation of percolating clusters and epidemic spreading.
In this Brief Report, we extend the previous BP formalism to consider a directed Boolean cascade model on a bipartite network composed of logic AND gates and OR gates. Then we study the cascading failure problem based on the model, by analyzing the avalanche size distribution in the Boolean cascade dynamics.
This generalization may be useful in further work for studying avalanche dynamics in bipartite or multitype networks in various systems and may provide a guideline for constructing a formalism of the BP for interacting networks \cite{leicht,son,brummitt}.

{\it Boolean cascade model---} The Boolean cascade model is defined on a directed bipartite network. The network is composed of two types of nodes: Boolean AND gates and Boolean OR gates. Each of these nodes is connected by a directed edge to a node of the other type, thus defining a bipartite network. This model was developed in previous studies on the basis of the reaction blockade cascade in metabolic networks \cite{lemke,cmghim}: When a reaction is blocked by, e.g., the knockout of the gene(s) catalyzing it, its product cannot be produced, which in turn blocks other reactions,  i.e., an avalanche occurs. In the dynamics, the metabolites act as an OR gate, since any one of the reactions producing that metabolite can make it turn on. In contrast, the reaction node is represented by the AND gate, since all the input metabolites should be present for the reaction to be activated. 

\begin{figure}
\includegraphics[width=8.5cm]{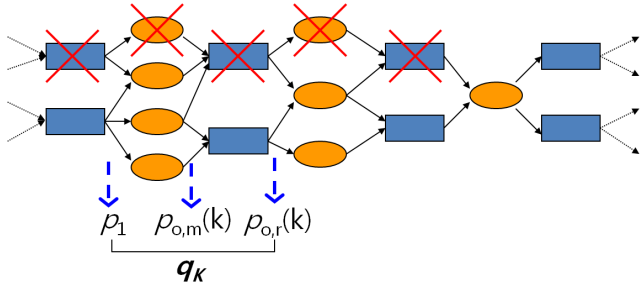}
\caption{(Color online) Schematic illustration of  avalanche dynamics in a Boolean bipartite network.
When a reaction (rectangle) is deactivated, a metabolite (ellipse) would be deactivated if all the reactions
have that metabolite as a product.
A reaction cannot take place if any of the substrate metabolites is absent. Thus, metabolites and
reactions correspond to logic OR and AND gates, respectively. When a metabolite is generated from a single reaction
(with probability $p_1$), it can block $\ell$ reactions (with  probability $p_{o,m}(\ell)$) and each of the
metabolite branches $k_i$ ($i=1,\dots, \ell$)  (with  probability $p_{o,r}(k_i)$).
The resulting value $K=k_1+\cdots+k_{\ell}$ corresponds to the branching number, which has probability $q_K$, as derived in Eq. (1).}
\label{fig1}
\end{figure}

Specifically, the dynamic rule of the Boolean cascade model is as follows, which is also schematically illustrated in Fig.~\ref{fig1}:
\begin{enumerate}
\item[(D1)] Initially, all nodes in both subsets (metabolites and reactions) are active, that is, their Boolean states are set to $b_i=1$ for all nodes $i=1,\dots, N$. Then a single reaction node, say $r_{\textrm{init}}$, is turned off, that is, $b_{r_{\textrm{init}}}=0$.
\item[(D2)] For all the metabolite nodes connected with the newly inactivated reaction nodes, update its Boolean state according to the Boolean OR function, $b_m=b_{r_1}\lor b_{r_2}\lor\cdots\lor b_{r_{k_{i,m}}}$, where $r_i$ are the reactions producing the metabolite $m$.  
\item[(D3)] For all the reaction nodes connected with the newly inactivated metabolite nodes, update its Boolean state according to the Boolean AND function, $b_r=b_{m_1}\land b_{m_2}\land\cdots\land b_{m_{k_{i,r}}}$, where $m_i$ are the metabolites producing the reaction $r$.  
\item[(D4)] Repeat (D2)--(D3) until no more inactivation occurs, and  
the total number of inactivated metabolite nodes, called the avalanche size, is recorded. 
\item[(D5)] Repeat the above procedure for each starting reaction, and obtain the avalanche size distribution $p_a(s)$.
\end{enumerate}

{\it Branching process approach---} 
The propagation of the Boolean cascade can be understood in view of the branching process.
This approach is based on the assumption that the cascade does not form a significant fraction of loops
during propagation. This amounts to the assumption that the metabolite node should be of in-degree
$1$ to be deactivated.
To start with, we assume for simplicity that the in-degree and out-degree of a node are uncorrelated, that is,
$p(j,k)=p_{i}(j)p_{o}(k)$, where $p(j,k)$ is the joint probability distribution of the in-degree $j$
and the out-degree $k$.

Next, we reduce the bipartite network to a unipartite network composed only of
reactions. In this transformation, the probability distribution $q_K$ of the effective branching ratio $K$
for a given reaction is derived. This can be more easily obtained in terms of the generating function,
\begin{align}
{\cal Q}&(\omega) \equiv \sum_{K=0}^{\infty}q_K\omega^{K}\nonumber\\
& = (1-p_1) + p_1\sum_{\ell=0}^{\infty}\sum_{k_1,\cdots,k_{\ell}=0}^{\infty}p_{o,m}(\ell)\prod_{i=1}^{\ell}p_{o,r}(k_i)\omega^{\sum_{i=1}^{\ell}k_i} \nonumber \\
&= (1-p_1) + p_1\sum_{\ell=0}^{\infty}p_{o,m}(\ell)Q_r(\omega)^{\ell}\nonumber\\
&=(1-p_1) + p_1Q_m(Q_r(\omega)),
\end{align}
where $p_1$ is the probability that the in-degree of a metabolite encountered is one (see Fig.~1),
and the $k_i$'s and $\ell$ are randomly chosen from the probability distributions $p_{o,r}(k)$ and $p_{o,m}(k)$, respectively.
Then, the effective branching ratio $K$ is determined by $K=k_1+k_2+\cdots+k_{\ell}$. $Q_m(\omega)$ and
$Q_r(\omega)$ are the generating functions defined as
\begin{equation}
{Q}_m(\omega)=\sum_{k=0}^{\infty}p_{o,m}(k)\omega^k~~{\quad \rm and \quad}~~
{Q}_r(\omega)=\sum_{k=0}^{\infty}p_{o,r}(k)\omega^k.
\end{equation}
The criticality condition is $\langle K \rangle \equiv{\cal Q}'(\omega)|_{\omega=1}=1$.

Next, we solve for the avalanche size distribution $p_a(s)$ of the Boolean dynamics with the assumption that
it proceeds in a branching tree pattern. Using the standard steps of the BP approach, we write
the avalanche size distribution $p_a(s)$ as
\begin{equation}
p_a(s) =\sum_{K=0}^{\infty} q_K \sum_{s_1=1}^{\infty}\cdots\sum_{s_K=1}^{\infty}
p_a(s_1)\cdots p_a(s_K)\delta_{{\sum_{i=1}^K s_i},s-1},
\end{equation}
where a seed deactivated reaction has $K$ branches, and from each of the branches, $s_i$-sized successive deactivated reactions
follow under the condition $\sum_i^K s_i=s-1$.
This formula can be rewritten in terms of the generating functions ${\cal P}_a(z)=\sum_{s=1}^{\infty} p_a(s)z^s$
and ${\cal Q}(\omega)$ as
\begin{equation}
{\cal P}_a(z)=z{\cal Q}({\cal P}_a(z)).
\label{self}
\end{equation}
Assuming that the avalanche size distribution follows a power law, i.e., $p_a(s)\sim s^{-\tau}$,
we can expand the generating function  near $z=1$ %similar to the case of ~Eq. (\ref{generating})
as
\begin{equation}
{\cal P}_a(z) \simeq 1 - (1-z) + c_1(1-z)^{\tau-1}+ c_2(1-z)^2+\cdots.
\end{equation}

Let us consider the criticality condition ${\cal Q}'(\omega)|_{\omega=1}=1$.
Applying it to Eq.~(1), we obtain the mean branching ratio as
\begin{equation}
{\cal Q}'(\omega)|_{\omega=1} = p_1 Q'_m(Q_r(1))Q'_r(1) = p_1\langle k_{o,m}\rangle\langle k_{o,r}\rangle,
\end{equation}
where $\langle k_{o,m}\rangle$ and  $\langle k_{o,r}\rangle$ are the mean out-degrees of the metabolites
and the reactions, respectively. Therefore, we have to tune the parameter $p_1$ to maintain the criticality condition.
$p_1$ is the probability of reaching a node (metabolite) whose inward degree is one, which is given by
the formula $p_k=k p_{i,m}(k)/\langle k_{i,m} \rangle$ with $k=1$ as $p_1=p_{i,m}(1)/\langle{k_{i,m}}\rangle$, so that
the mean branching ratio becomes
\begin{equation}
\langle K \rangle = \frac{p_{i,m}(1)\langle k_{o,m} \rangle\langle{k_{o,r}}\rangle}{\langle{k_{i,m}}\rangle}.
\label{bratio}
\end{equation}

{\it Power-law degree distributions---}
As an example, we consider the cases where $p_{o,m}(k)$ and $p_{o,r}(k)$ follow power laws, i.e.,
$p_{o,m}(k)\sim k^{-\gamma_m}$ and $p_{o,r}(k)\sim k^{-\gamma_r}$.
As shown below, this is an interesting regime not only because we have nontrivial scaling behavior in $p_a(s)$, but also being relevant for real-world metabolic networks \cite{jeong}.

In this case, we can determine the expansion of ${\cal Q}(\omega)$ near $\omega\approx1$ from the smaller of $\gamma_m$ and $\gamma_r$, i.e., $\gamma=\min[\gamma_m,\gamma_r]$, as
\begin{equation}
{\cal Q}(\omega) \simeq 1 - (1-\omega) +
\left\{\begin{array}{ll}
A_1(1-\omega)^{\gamma-1} & (2<\gamma<3), \\
-A_2(1-\omega)^2\ln(1-\omega) & (\gamma=3),\\
A_3(1-\omega)^2 & (\gamma>3). \\
\end{array}\right.
\label{generating}
\end{equation}
To solve the self-consistent equation~(\ref{self}), we set $\omega={\cal P}_a(z)$, and consequently,
$z={\cal P}_a^{-1}=\omega/{\cal Q}(\omega)\approx 1-A_1(1-\omega)^{\gamma-1}+\cdots$.
Thus, we obtain 
\begin{equation}
\tau = 
\left\{\begin{array}{ll}
\gamma/(\gamma-1) & (2<\gamma\le 3), \\
3/2 & (\gamma>3). \\
\end{array}\right.
\end{equation}

For the case in which all $p_{i,m}(k)$, $p_{o,m}(k)$, and $p_{o,r}(k)$ follow the same power-law distribution
with the same exponent  $\gamma$, $\langle K \rangle=\zeta(\gamma-1)/\zeta(\gamma)^2$, which is larger
than $1$ for finite $\gamma$. Thus, the BP shows a supercritical behavior.
In the supercritical regime $\langle K \rangle > 1$~\cite{dslee}, the avalanche size distribution behaves as
\begin{equation}
p_a(s) \sim \left\{\begin{array}{ll}
s^{-\gamma/(\gamma-1)} & (s\ll s_c), \\
s^{-3/2}e^{-s/s_c} & (s\gg s_c), \\
\end{array}\right.
\end{equation}
in which $s_c \sim |\langle K \rangle-1|^{-\alpha}$ with  $\alpha=(\gamma-1)/(\gamma-2)$
for $2<\gamma<3$, and
\begin{equation}
p_a(s) \sim s^{-3/2}e^{-s/s_c}
\end{equation}
with $s_c \sim |\langle K \rangle-1|^{-2}$ for $\gamma>3$.

{\it Numerical simulations---}
To test the BP predictions with simulation results, we first construct a model Boolean bipartite network, which is a generalization of configuration model, as follows:
\begin{enumerate}
\item[(S1)] The system is composed of $N$ nodes, which are composed of two subsets, $\mathcal{M}$ and $\mathcal{R}$,
of equal size, $N/2$.
\item[(S2)] For a node $v\in \mathcal{M}$, assign in-degree $k_{i,v}$ and out-degree $k_{o,v}$ randomly from the
probability distributions $p_{i,m}(k)$ and  $p_{o,m}(k)$, respectively.
Similarly, for a node $w\in \mathcal{R}$, assign in-degree $k_{i,w}$ and out-degree $k_{o,w}$ from the
probability distributions $p_{i,r}(k)$ and $p_{o,r}(k)$, respectively.
\item[(S3)] Choose an ordered pair of vertices $(v,w)$ and join them by the directed edge $v\to w$
if they belong to different subsets, provided that there remain a free outward arrow at $v$ and a free inward arrow at $w$ and that the pair is not connected yet, that is, we disallow multiple arrows between the node pair.
\item[(S4)] Repeat (S3) until there is no stub left.
\item[(S5)] Assign Boolean OR gates to all nodes in the subset ${\cal M}$
and assign Boolean AND gates to those in ${\cal R}$.
\end{enumerate}

\begin{figure}
\includegraphics[width=.9\linewidth]{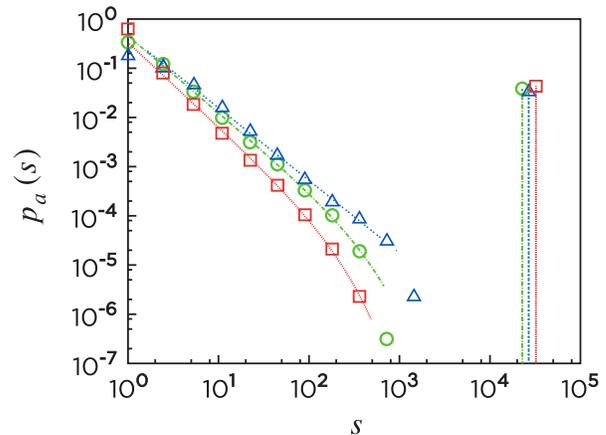}
\caption{(Color online) The avalanche size distribution $p_a(s)$ for the Boolean cascade model
on the directed bipartite scale-free network with the in- and out-degree
distribution $p_d(k)\sim k^{-\gamma}$ and $\gamma=2.5$ (red), $3.5$ (green),
and $4.5$ (blue). The curves are a good fit to the theoretical function
$p_a(s)=As^{-\tau}\exp(-s/s_c)$ with  $\tau$ values of  $1.71$ (red), $1.56$ (green), and $1.47$ (blue). Note that we are in the supercritical regime so that isolated peaks at large $s$ appear.
The system size is $N=10^5$.
}
\end{figure}

Simulation results for the Boolean cascade dynamics on the model networks with in- and out-degree distribution of same power-law form, with $\gamma=2.5$, $3.5$, and $4.5$ are shown in Fig.~2, showing a good agreement with the BP prediction.

{\it Real-world metabolic networks---}
We now consider the Boolean cascade dynamics on the real-world metabolic networks of {\it Escherichia coli} and {\it Saccharomyces cerevisiae} \cite{palsson},
which contain 1,188 metabolites and 1,489 reactions
for {\it E.~coli} and 680 metabolites and 852 reactions for {\it S.~cerevisiae}, respectively.
The out-degree distributions for the metabolites and the reactions in the metabolic networks are measured. For both networks, the out-degree distribution for metabolites is found to follow approximately a power law with the exponent of  $\gamma_{o,m}\approx 2.2$ (Fig.~3, insets), in agreement with previous results \cite{jeong}. 
On the other hand, the out-degree distribution for reactions is found to decay exponentially (Fig.~3, insets).

\begin{figure}[t]
\includegraphics[width=.9\linewidth]{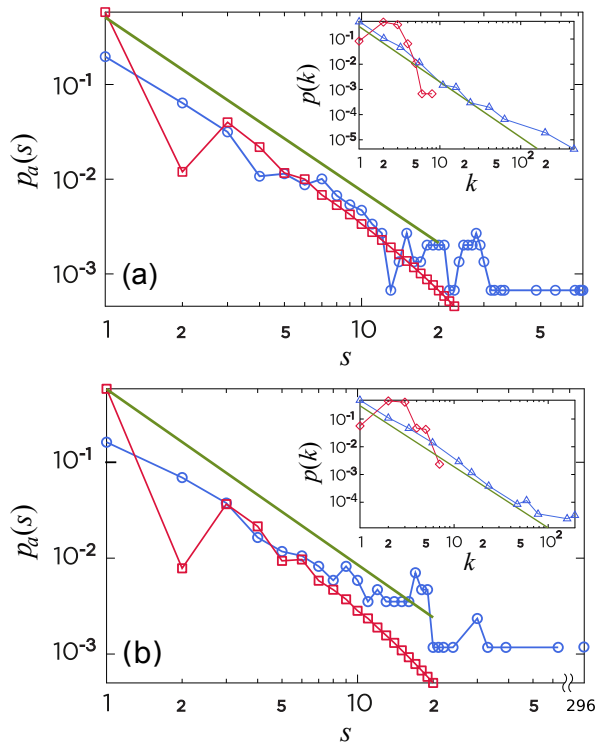}
\caption{(Color online) 
	The avalanche size distributions on real-world metabolic networks.
	(a) is for {\it E. coli} and (b) is for {\it S. cerevisiae}.
	Circles are for simulation results, and squares are for BP predictions with empirical degree distributions.
	Solid lines are guidelines with the slope of $-1.83$ which satisfies the relation $\tau=\gamma/(\gamma-1)$ with $\gamma=2.2$.
	Insets are out-degree distributions of metabolites (triangles) and reactions (diamonds).
	Solid guidelines in the insets have a slope $-2.2$.
}\label{fig3}
\end{figure}

We run the Boolean cascade dynamics on these networks and measure the avalanche size distribution $p_a(s)$. 
From numerical simulation, we found that the avalanche size distribution for both species has an approximate power-law tail, with the exponent $\tau\approx1.8$ (Fig.~3). 
This power-law exponent is not far from the BP prediction by Eq.~(9) with $\gamma=2.2$.

To further compare the numerical simulation results with BP predictions, we numerically solved Eq.~(4) by plugging in the empirical degree distributions, the result of which is depicted also in Fig.~3. 
The theoretical $p_a(s)$ deviates from a pure power law, because the dynamics is super-critical, as $\langle K\rangle\approx 3.26$ ({\it E.~coli}) and $3.85$ ({\it S.~cerevisiae}), respectively.
Also, real-world metabolic networks contain the degree-degree correlation and clustering, as well as are finite, which lead to deviations from the mean-field type analytic predictions \cite{gleeson}.
For example, we note the peaks in the empirical $p_a(s)$ near $s\approx20\sim30$, that are absent in theoretical $p_a(s)$, which might arise due to the existence of compact modules or cycles in the metabolic networks, such as TCA cycle.
Despite these complications, the power-law behaviors obtained from numerical simulations and BP calculations agree reasonably, demonstrating the usefulness of BP approach to understanding the scaling behaviors.

{\it Summary---}
In conclusion, we have studied the Boolean cascade dynamics occurring in directed bipartite networks of Boolean AND and OR logic gates, inspired by
the cascading failure of reactions in metabolic networks. 
A branching process approach was developed to study this type of dynamics. 
Theoretically obtained exponent for the avalanche size distribution is in good agreement with the simulation results for the model network as well as for two empirical microbial metabolic networks.  \\

\acknowledgments
BK would like to thank Y. W. Kim for helpful discussions.
This work was supported by NRF research grants funded by MEST
[Nos.~2010-0015066 (BK) and 2011-0014191 (K-IG)].

%%%%%%%%%%%%%%%%%%%%%%%%%%%%%%%%%%%%%%%%%%%%%%%%%%%%%%%%%%%%%%%%%%%%%%%%%%%%%%
% Bibliography
%%%%%%%%%%%%%%%%%%%%%%%%%%%%%%%%%%%%%%%%%%%%%%%%%%%%%%%%%%%%%%%%%%%%%%%%%%%%%%

\end{document}